\documentclass{rnaastex}

\begin{document}

\title{Identification of SIPS~J2045$-$6332 as a partially resolved binary}

\correspondingauthor{Niall R Deacon}
\email{deacon@mpia.de}

\author{Niall R. Deacon}
\affiliation{MPIA, Konigstuhl, Heidelberg, Germany}
\affiliation{Centre for Astrophysics Research, University of Hertfordshire, College Lane, Hatfield, UK}

\author{Kyle L. Sobanja}
\affiliation{Centre for Astrophysics Research, University of Hertfordshire, College Lane, Hatfield, UK}

\author{Leigh C. Smith}
\affiliation{Institute of Astronomy, University of Cambridge, Madingley Road, Cambridge, UK}
\affiliation{Centre for Astrophysics Research, University of Hertfordshire, College Lane, Hatfield, UK}
\keywords{binaries: visual, brown dwarfs}


\section{Introduction} 
\cite{Hoekstra2005} intorduced a method to identify blended visual binary stars using image shape measurement techniques developed for cosmological weak lensing surveys. This has been applied to find binary stars and brown dwarfs using PTF \citep{Terziev2013} and Pan-STARRS\,1 \citep{Deacon2017}. See these three papers for a detailed description of the methods used and parameters measured. We used this technique to measure the image shapes of a sample of brown dwarfs drawn from the list maintained by J Gagn\'{e} \footnote{\url{https://docs.google.com/spreadsheets/d/1shqSyDMEForWLrVTqYEHTt8T_TDqwCTX00ZwIvP97oU}} on images taken from the VISTA Hemisphere Survey (VHS; \citealt{McMahon2013}). We identified SIPS~J2045-6332 as being a possible binary as it had a total ellipticity greater than 0.05. 

SIPS~J2045-6332 \citep{Deacon2007} is an ultracool dwarf at $d=25.0\pm^{2.5}_{2.2}$\,pc \citep{Marocco2013} classified as an M9 by \cite{Reid2008} using an optical spectrum and as an L1 from an infrared spectrum by \cite{Marocco2013}. \cite{Marocco2013} also noted that this object showed spectral peculiarities which suggested it had an unresolved cooler companion.

\section{Results} 
We measured the shape parameters of SIPS~J2045-6332 across 6 VHS images, three each in $J$ and $K$. The image shape measurement technique we use requires a window function to suppress noise at large distances from the target's photocentre. \cite{Deacon2017} used a gaussian with a FWHM equal to that of the measured seeing for each image. We found that this worked well for images with seeing $\ge$0.7'' but resulted in anomalously low ellipticity measurements for images with better seeing. This is likely due to much of the companion's light being suppressed by the window function in these images. Hence we changed our window function to a gaussian with a FWHM equal to double that of the measured seeing. The shape measurement results are shown in Table~\ref{tab1}. 

\section{Discussion} 
We find $e_{1,cor}$ measurements that are always positive and larger in magnitude than the $e_{2,cor}$ values which are always negative. The $e_1$ ellipticity represents a distortion in the $+$ polarisation with a positive value indicating a distortion along the E-W axis (or the $\rightarrow$ direction). The $e_2$ ellipticity represents a distortion in the $\times$ polarisation with a negative value representing a distortion in the $\searrow$ direction. As North is down and East left on the VHS images used, this implies that the binary is tilted towards the NW-SE direction. The combination of these individual measurements show that this binary likely has a position angle of around 290$^{\circ}$ East of North.

The measured ellipticity is always higher for $K$-band images than $J$-band images. This implies that the flux ratio between the two components is closer to unity in $K$ than in $J$. \cite{Marocco2013} suggested a combination of an L1 and a T5 produced the best match for SIPS~J2045-6332's spectrum but noted that their constraint on the secondary spectral type was weak. Our results suggest that the secondary is more likely to be a late-L dwarf given the implied $J-K$ colour.

\section{Conclusions}
We have shown that SIPS~J2045-6332, already a candidate spectral mix binary, is a candidate partially resolved binary. The individual ellipicity measurements suggest the secondary component is likely to be a late L dwarf. This object would be an excellent target for high resolution imaging. The shape measurement technique used here provides an excellent way to screen objects for binarity using only survey data. This can be used to both detect binaries and to remove them from samples where they would be contaminants. Image shape measurement techniques will be useful for identifying binaries using future surveys such as those undertaken with LSST.

\begin{deluxetable}{lcrrrrrrc}
\tablecaption{\label{tab1}}
\tablehead{
\colhead{Image} & \colhead{Filter} & \colhead{$e_1$}& \colhead{$e_{1,cor}$}& \colhead{$e_2$}& \colhead{$e_{2,cor}$}& \colhead{$e$}& \colhead{$e_{cor}$}& \colhead{Seeing ('')}
}
\startdata
v20130513\_00475&$K$&0.163&0.173&$-$0.134&$-$0.113&0.211&0.207&0.73\\
v20130513\_00487&$J$&0.103&0.111&$-$0.107&$-$0.070&0.145&0.131&0.78\\
v20130519\_00837&$K$&0.279&0.262&$-$0.128&$-$0.110&0.308&0.284&0.60\\
v20130519\_00841&$K$&0.258&0.235&$-$0.152&$-$0.117&0.299&0.263&0.67\\
v20130519\_00849&$J$&0.136&0.124&$-$0.102&$-$0.066&0.170&0.140&0.78\\
v20130519\_00853&$J$&0.176&0.154&$-$0.098&$-$0.083&0.201&0.175&0.66\\
\enddata
\tablecomments{A table showing the individual ellipicity measurements for each image. Those columns marked $_{cor}$ are the measurements after correction for telescope-induced PSF anisotropy.. The uncertainty on each measurement from photon noise will be approximately 0.001 although the uncertainty introduced by the PSF anisotropy correction will likely be larger than this.}
\end{deluxetable}  

\acknowledgments
We thank Mike Irwin for helping us to access the VHS data. K.L.S. was supported by a Nuffield Research Placement and Setpoint Hertfordshire. Based on data products from observations made with ESO Telescopes at the La Silla or Paranal Observatories under ESO programme ID 179.A-2010.
\bibliography{ndeacon}

\begin{thebibliography}{7}
\expandafter\ifx\csname natexlab\endcsname\relax\def\natexlab#1{#1}\fi

\bibitem[{Deacon \& Hambly(2007)}]{Deacon2007}
Deacon, N.~R., \& Hambly, N.~C. 2007, Astron. Astrophys., 468, 163

\bibitem[{Deacon {et~al.}(2017)Deacon, Magnier, Best, Liu, Dupuy, Chambers,
  Draper, Flewelling, Metcalfe, Tonry, Wainscoat, \& Waters}]{Deacon2017}
Deacon, N.~R., {et~al.} 2017, Mon. Not. R. Astron. Soc., 468, 3499

\bibitem[{Hoekstra {et~al.}(2005)Hoekstra, Wu, \& Udalski}]{Hoekstra2005}
Hoekstra, H., Wu, Y., \& Udalski, A. 2005, ApJ, 626, 1070

\bibitem[{Marocco {et~al.}(2013)Marocco, Andrei, Smart, Jones, Pinfield,
  Day-Jones, Clarke, Sozzetti, Lucas, Bucciarelli, \& Penna}]{Marocco2013}
Marocco, F., {et~al.} 2013, Astron. J., 146, 161

\bibitem[{McMahon {et~al.}(2013)McMahon, Banerji, Gonzalez, Koposov, Bejar,
  Lodieu, \& Rebolo}]{McMahon2013}
McMahon, R.~G., Banerji, M., Gonzalez, E., Koposov, S.~E., Bejar, V.~J.,
  Lodieu, N., \& Rebolo, R. 2013, The Messenger, 154, 35

\bibitem[{Reid {et~al.}(2008)Reid, Cruz, Kirkpatrick, Allen, Mungall, Liebert,
  Lowrance, \& Sweet}]{Reid2008}
Reid, I., Cruz, K.~L., Kirkpatrick, J.~D., Allen, P.~R., Mungall, F., Liebert,
  J., Lowrance, P., \& Sweet, A. 2008, Astron. J., 136, 1290

\bibitem[{Terziev {et~al.}(2013)Terziev, Law, Arcavi, Baranec, Bloom, Bui,
  Burse, Chorida, Das, Dekany, Kraus, Kulkarni, Nugent, Ofek, Punnadi,
  Ramaprakash, Riddle, \& Tendulkar}]{Terziev2013}
Terziev, E., {et~al.} 2013, ApJS, 206, 18

\end{thebibliography}
\bibliographystyle{apj}
\end{document}